\documentclass[preprint,aps,prd,amsmath,amssymb,showpacs,nofootinbib]{revtex4}
\usepackage{graphicx}
\usepackage{bm}

\newcommand{\beq}{\begin{equation}}
\newcommand{\eeq}{\end{equation}}
\newcommand{\bea}{\begin{eqnarray}}
\newcommand{\eea}{\end{eqnarray}}
\newcommand{\non}{\nonumber}
\newcommand{\Non}{\nonumber\\}
\newcommand{\Ref}[1]{(\ref{#1})}
\newcommand{\sss}{static spherically symmetric\ }

\begin{document}

\title{Scalar wormholes with nonminimal derivative coupling}

\author{Sergey V. Sushkov$^{1,2,}$}
\email{sergey_sushkov@mail.ru}
\affiliation{$^1$ Department of Mathematics and Department of Physics,
Kazan Federal University, Kremlevskaya str. 18, Kazan 420008, Russia\\
$^2$ Physics Department, CSU Fresno, Fresno, CA 93740-8031}

\author{Roman Korolev}
\affiliation{Department of Physics,
Kazan Federal University, Kremlevskaya str. 18, Kazan 420008, Russia}

\begin{abstract}
We consider static spherically symmetric wormhole configurations in a gravitational theory of a scalar field with a potential $V(\phi)$ and nonminimal derivative coupling to the curvature describing by the term  $(\varepsilon g_{\mu\nu} + \kappa G_{\mu\nu} ) \phi^{,\mu}\phi^{,\nu}$ in the action. We show that the flare-out conditions providing the geometry of a wormhole throat could fulfilled both if $\epsilon=-1$ (phantom scalar) and $\epsilon=+1$ (ordinary scalar). Supposing additionally a traversability, we construct numerical solutions describing traversable wormholes in the model with arbitrary $\kappa$, $\epsilon=-1$ and $V(\phi)=0$ (no potential). The traversability assumes that the wormhole possesses two asymptotically flat regions with corresponding Schwarzschild masses. We find that asymptotical masses of a wormhole with nonminimal derivative coupling could be positive and/or negative depending on $\kappa$. In particular, both masses are positive only provided $\kappa<\kappa_1\le0$, otherwise one or both wormhole masses are negative. In conclusion, we give qualitative arguments that a wormhole configuration with positive masses could be stable.
\end{abstract}

 \maketitle

\section{Introduction}
Nonminimal generalizations of general relativity imply a straightforward coupling between matter fields and the spacetime curvature and play an important role in modern theoretical physics. The well-known example of nonminimal theories could be represented by scalar-tensor theories of gravity with the action generally given as\footnote{Throughout this paper we use units such that $G=c=1$. The metric signature is $(- + + +)$ and the conventions for curvature tensors are $R^\alpha_{\beta\gamma\delta} = \Gamma^\alpha_{\beta\delta,\gamma} - ...$ and $ R_{\mu\nu} = R^\alpha_{\mu\alpha\nu}$.}
\beq\label{STT}
  S=\int dx^4\sqrt{-g}\big[F(\phi,R)+K(\phi,X)+V(\phi)\big]+S_m,
\eeq
where $\phi$ is the scalar field, $X=\phi_\mu\phi^\mu$, $R$ is the scalar curvature, and ${S}_m$ is an action of ordinary matter (not including the scalar field). Here the function $F(\phi,R)$ provides a nonminimal coupling between the scalar field $\phi$ and the curvature, $K(\phi,X)$ represents a generalized kinetic term, and $V(\phi)$ is a scalar field potential. Note that the theory \Ref{STT} includes a lot of extensively investigated models, among them the $f(R)$ gravity and the Gauss-Bonnet gravity, the K-essence scalar theory, models with quintessence, quintom, phantom, dilaton, tachyon, and so on.\footnote{The complete list of references concerning various applications and aspects of scalar-tensor theories would include hundreds of items, and its survey goes beyond the present paper. For review see, for example, Ref. \cite{FM}.}

A further extension of scalar-tensor theories is represented by models with nonminimal couplings between derivatives of the scalar field and the curvature. In general, one could have various forms of such couplings. For instance in the case of four derivatives one could have the terms $\kappa_1 R\phi_{,\mu}\phi^{,\mu}$, $\kappa_2 R_{\mu\nu}\phi^{,\mu}\phi^{,\nu}$, $\kappa_3 R \phi\square\phi$, $\kappa_4 R_{\mu\nu} \phi\phi^{;\mu\nu}$, $\kappa_5 R_{;\mu} \phi\phi^{,\mu}$ and $\kappa_6 \square R \phi^2$, where the coefficients $\kappa_1,\dots,\kappa_6$ are coupling parameters with dimensions of length-squared. However, as it was discussed in \cite{Ame,Cap:99,Cap:00,Sus}, using total divergencies and without loss of generality one can keep only the first two terms.

As was shown by Amendola \cite{Ame}, a theory with derivative couplings cannot be recasting into Einsteinian form by a conformal rescaling $\tilde g_{\mu\nu}=e^{2\omega}g_{\mu\nu}$. He also supposed that an effective cosmological constant, and then the inflationary phase can be recovered without considering any effective potential if a nonminimal derivative coupling is introduced. Amendola himself \cite{Ame} investigated a cosmological model with the only derivative coupling term $\kappa_2 R_{\mu\nu}\phi^{,\mu}\phi^{,\nu}$ and presented some analytical inflationary solutions. A general model containing $\kappa_1 R\phi_{,\mu}\phi^{,\mu}$ and $\kappa_2 R_{\mu\nu}\phi^{,\mu}\phi^{,\nu}$ has been discussed by Capozziello et al. \cite{Cap:99,Cap:00}. They showed that the de Sitter spacetime is an attractor solution in the model.

Note that generally field equations in the model with terms $\kappa_1 R\phi_{,\mu}\phi^{,\mu}$ and $\kappa_2 R_{\mu\nu}\phi^{,\mu}\phi^{,\nu}$ contain higher (third) derivatives of the metric and the scalar field. However, as was shown in our work \cite{Sus}, the order of field equations reduces to second one in the particular case when the kinetic term is only coupled to the Einstein tensor, i.e., $\kappa G_{\mu\nu}\phi^{,\mu}\phi^{,\nu}$. In Refs. \cite{Sus,SarSus} we studied in detail exact cosmological scenarios with a nonminimal derivative coupling $\kappa G_{\mu\nu}\phi^{,\mu}\phi^{,\nu}$, examining both the quintessence and the phantom cases with zero and constant potentials. It is worth noticing that taking into account the nonminimal derivative coupling reveals new interesting features in a cosmological behavior. In general, we found \cite{Sus,SarSus} that the universe has two quasi-de Sitter phases and transits from one to another without any fine-tuned potential, determined only by the coupling parameter. Further investigations of cosmological models with nonminimal derivative coupling have been continued in Refs. \cite{Gao,Granda_10,Granda_11a,Granda_11b,Granda_11c,Granda_11d,Granda_11e, Sadjadi,Banijamali,Gubitosi}.\footnote{It is also worth mentioning a series of papers devoted to a nonminimal modification of the Einstein-Yang-Mills-Higgs theory \cite{BalDehZay:07} (see also a review \cite{BalDehZay} and references therein).}

In this paper we will study static spherically symmetric wormhole configurations in the scalar-tensor theory with the nonminimal derivative coupling. The wormholes supported by a minimally coupled scalar field are well known in the literature (see, for example, Refs. \cite{Ellis,Bro73,SusZhang}). It is also known that they have a number of features unacceptable with the physical point of view. In particular, they have negative asymptotical Schwarzschild masses and are unstable \cite{SusZhang,shin02,gon08,Bro_etal_11}. We will show that the model with non-minimal derivative coupling provides scalar wormholes of a new type, possessing positive asymptotical masses, and give some arguments on their stability.

\section{Model with non-minimal derivative coupling\label{II}}

\subsection{Action and field equations}

Let us consider a gravitational theory of a scalar field $\phi$ with nonminimal derivative coupling to the curvature which is described by the following action:
\begin{equation}\label{action}
S=\int d^4x\sqrt{-g}\left\{ \frac{R}{8\pi} -\big[\varepsilon
g_{\mu\nu} + \kappa G_{\mu\nu} \big] \phi^{,\mu}\phi^{,\nu} -2
V(\phi)\right\},
\end{equation}
where $V(\phi)$ is a scalar field potential, $g_{\mu\nu}$ is a
metric, $g=\det(g_{\mu\nu})$, $R$ is the scalar curvature,
$G_{\mu\nu}$ is the Einstein tensor, and $\kappa$ is the
derivative coupling parameter with the dimension of length-squared.

Varying the action \Ref{action} with respect to the metric
$g_{\mu\nu}$ leads to the gravitational field equations
\beq\label{eineq}
G_{\mu\nu}=8\pi\big[\varepsilon T_{\mu\nu}
+\kappa \Theta_{\mu\nu}\big]-8\pi g_{\mu\nu} V(\phi),
\eeq
with
\bea
\label{T}
T_{\mu\nu}&=&\nabla_\mu\phi\nabla_\nu\phi-
{\textstyle\frac12}g_{\mu\nu}(\nabla\phi)^2, \\
\Theta_{\mu\nu}&=&-{\textstyle\frac12}\nabla_\mu\phi\,\nabla_\nu\phi\,R
+2\nabla_\alpha\phi\,\nabla_{(\mu}\phi R^\alpha_{\nu)}
\nonumber\\
&&
+\nabla^\alpha\phi\,\nabla^\beta\phi\,R_{\mu\alpha\nu\beta}+\nabla_\mu\nabla^\alpha\phi\,\nabla_\nu\nabla_\alpha\phi
\nonumber\\
&&
-\nabla_\mu\nabla_\nu\phi\,\square\phi-{\textstyle\frac12}(\nabla\phi)^2
G_{\mu\nu}
\nonumber\\
&&
+g_{\mu\nu}\big[-{\textstyle\frac12}\nabla^\alpha\nabla^\beta\phi\,\nabla_\alpha\nabla_\beta\phi
+{\textstyle\frac12}(\square\phi)^2
\nonumber\\
&& -\nabla_\alpha\phi\,\nabla_\beta\phi\,R^{\alpha\beta}
\big]. \label{Theta}
\eea
Similarly, variation of the action
\Ref{action} with respect to $\phi$ provides the scalar field
equation of motion:
\beq \label{eqmo}
[\varepsilon
g^{\mu\nu}+\kappa G^{\mu\nu}]\nabla_{\mu}\nabla_\nu\phi=V_\phi,
\eeq
where $V_\phi\equiv dV(\phi)/d\phi$.

Note that due to Bianchi identity $G^\mu_{\nu;\mu}=0$ the right hand side of Eq. \Ref{eineq} should obey the relation
\beq
\label{BianchiT}
\big[(\varepsilon T^\mu_{\nu}
+\kappa \Theta^\mu_{\nu})-\delta^\mu_{\nu} V(\phi)\big]_{;\mu}=0.
\eeq
One can check straightforwardly that the substitution of expressions \Ref{T} and \Ref{Theta} into \Ref{BianchiT} yields the equation of motion of scalar field \Ref{eqmo}. Therefore, the relation \Ref{BianchiT} will take place provided Eq. \Ref{eqmo} is fulfilled. In other words, Eq. \Ref{eqmo} is a differential consequence of Eq. \Ref{eineq}.

\subsection{Field equations for a static spherically symmetric configuration}
Consider a \sss configuration in the theory \Ref{action}. In this case the spacetime metric can be taken as follows:
\begin{equation} \label{metric}
ds^2=-A(u)dt^2+A^{-1}(u)du^2+r^2(u)d\Omega^2,
\end{equation}
where $d\Omega^2=d\theta^2+\sin^2\theta d\varphi^2$ is the linear element of the unit sphere, and $A(u)$ and $r(u)$ are functions of the radial coordinate $u$. Also the scalar field $\phi$ depends only on $u$, so that $\phi=\phi(u)$.

Now the field equations \Ref{eineq} and \Ref{eqmo} yield
\begin{subequations} \label{fieldeq}
\bea
\label{00}
& \displaystyle
2\frac{r''}{r}+\frac{A'r'}{Ar}+\frac{r'^2}{r^2}-\frac{1}{Ar^2} = -4\pi\epsilon \phi'^2-8\pi A^{-1}V(\phi)
&
\Non
& \displaystyle
+8\pi\kappa \phi'^2\left( \frac{3A'r'}{2r} + \frac{Ar'^2}{2r^2}+\frac{1}{2r^2} +\frac{Ar''}{r}\right) +16\pi\kappa \phi'\phi''\frac{Ar'}{r},
&
\\
\label{11}
& \displaystyle
\frac{A'r'}{Ar}+\frac{r'^2}{r^2}-\frac{1}{Ar^2}  =
4\pi\epsilon\phi'^2-8\pi A^{-1}V(\phi) +8\pi\kappa\phi'^2\left(-\frac{1}{2r^2} +\frac{3Ar'^2}{2r^2}+\frac{3A'r'}{2r}\right),
&
\\
\label{22}
& \displaystyle
\frac12\frac{A''}{A}+\frac{r''}{r}+\frac{A'r'}{Ar}=-4\pi\epsilon\phi'^2-8\pi A^{-1}V(\phi)
&
\Non
& \displaystyle
+8\pi\kappa\phi'^2\left(\frac{A'r'}{r}+\frac12\frac{Ar''}{r}+\frac14\frac{A'^2}{A}+\frac14 A''\right) +8\pi\kappa\phi'\phi''\left(\frac{Ar'}{r}+\frac12 A'\right),
&
\\
\label{Eqmo}
& \displaystyle
\epsilon\phi''+\epsilon\phi'\left(\frac{A'}{A}+2\frac{r'}{r}\right) +\kappa\phi''\left(\frac{A'r'}{r}+\frac{Ar'^2}{r^2}-\frac{1}{r^2}\right)
&
\Non
& \displaystyle
+\kappa\phi'\left(\frac{A'r''}{r}-\frac{A'}{Ar^2}+3\frac{A'r'^2}{r^2}+\frac{A'^2r'}{Ar}  +2\frac{Ar'r''}{r^2}+\frac{A''r'}{r}\right)=A^{-1}V_{\phi},
&
\eea
\end{subequations}
where the prime means ${d}/{du}$, and Eqs. \Ref{00}, \Ref{11}, \Ref{22} are ${0\choose 0}$, ${1\choose 1}$, ${2\choose 2}$ components of Eq. \Ref{eineq}, respectively. Eqs. \Ref{fieldeq} represent a system of four ordinary differential equations of second order for three functions $r(u)$, $A(u)$, $\phi(u)$. As was mentioned above, Eq. \Ref{Eqmo} is a differential consequence of Eqs. \Ref{00}-\Ref{22}. It is also worth noticing that Eqs. \Ref{00}, \Ref{22} and \Ref{Eqmo} are of second order, while Eq. \Ref{11} is a first-order differential constraint for $r(u)$, $A(u)$, and $\phi(u)$.

Combining the above equations one can easily rewrite them into the more compact form:
\begin{subequations} \label{combfieldeq}
\bea
\label{00-11}
&& \displaystyle
\frac{r''}{r}=-4\pi\epsilon\phi'^2+4\pi\kappa A\left(\phi'^2\frac{r'}{r}\right)' +4\pi\kappa\phi'^2\frac{1}{r^2},
\\
\label{2(22)-(00)+(11)}
&& \displaystyle
(A'r^2)'=-16\pi r^2V+4\pi\kappa(AA'r^2\phi'^2)'+8\pi\kappa\phi'^2(AA'rr'+A^2r'^2-A),
\\
\label{(00)-(22)}
&& \displaystyle
A(r^2)''-A''r^2=2+4\pi\kappa[\phi'^2(2A^2rr'-AA'r^2)]'+8\pi\kappa A\phi'^2,
\\
\label{compacteqmo}
&& \displaystyle
\epsilon(Ar^2\phi')'+\kappa[\phi'(AA'rr'+A^2r'^2-A)]'=r^2 V_{\phi}.
\eea
\end{subequations}


\section{Wormhole solution \label{Sec}}
In this section we will focus our attention on wormhole solutions of the field equations \Ref{fieldeq}, or equivalently \Ref{combfieldeq}, obtained in the previous section. Note that Eqs. \Ref{fieldeq} are a rather complicated system of nonlinear ordinary differential equations, and we do not know if it is possible to find any exact analytical solutions to this system. Instead, we will construct wormhole solutions numerically studying previously their asymptotical properties near and far from the wormhole throat.

To describe a traversable wormhole the metric \Ref{metric} should possess a number of specific properties. In particular, (i) the radial coordinate $u$ runs through the domain $(-\infty,+\infty)$; (ii) there exist two asymptotically flat regions ${\cal R}_\pm:\ u\to\pm\infty$ connected by the throat; (iii) $r(u)$ has a global positive minimum at the wormhole throat $u=u_0$; without loss of generality one can set $u_0=0$, so that $r_0=\min\{r(u)\}=r(0)$ is the throat radius; (iv) $A(u)$ is everywhere positive and regular, i.e. there are no event horizons and singularities in the spacetime. Taking into account necessary conditions for the minimum of function, we obtain also
\beq\label{flareout}
r'_0=0, \quad r''_0>0,
\eeq
where the subscript `0' means that values are calculated at the throat $u=0$. Concerning wormholes, the above relations are known as the flare-out conditions.

\subsection{Initial condition analysis}
Let us consider the field equations at the throat $u=0$. By assuming $r'_0=0$, Eqs. \Ref{00} and \Ref{11} after a little algebra yield
\bea
\label{r0}
\displaystyle
\frac{1}{r_0^2} &=& -\frac{4\pi(\epsilon A_0\phi'^2_0-2V_0)}{1-4\pi\kappa A_0\phi'^2_0},
\\
\label{d2r0}
\displaystyle
\frac{r''_0}{r_0} &=& -\frac{4\pi\phi'^2_0(\epsilon-8\pi\kappa V_0)}{(1-4\pi\kappa A_0\phi'^2_0)^2}.
\eea
To provide the flare-out conditions \Ref{flareout} the right hand sides of Eqs. \Ref{r0}, \Ref{d2r0} should be positive. This is possible if $\phi'^2_0\not=0$ and the following inequalities take place:
\beq\label{ineq1}
\frac{\epsilon A_0\phi'^2_0-2V_0}{1-4\pi\kappa A_0\phi'^2_0}<0,
\eeq
\beq\label{ineq2}
\epsilon-8\pi\kappa V_0<0.
\eeq
For given $\epsilon$ and $\kappa$ these inequalities give restrictions for initial values of $A_0\phi'^2_0$ and $V_0=V(\phi_0)$ at the throat. Let us consider separately various cases.

\vskip6pt{\em 1.~$\kappa=0$.}~This is the case of a minimally coupled scalar field. Now \Ref{ineq2} yields $\epsilon<0$, i.e. $\epsilon=-1$, for any $V_0$. In turn, if $\epsilon=-1$ Eq. \Ref{ineq1} is fulfilled provided $V_0>-\frac12 A_0\phi'^2_0$. Thus, we have obtained a well-known result that a solution with the throat in general relativity with a minimally coupled scalar field is permitted only for phantom fields with negative kinetic energy (see, for example, \cite{Bro01}).

Further we will consider cases with nonminimal derivative coupling.
\vskip6pt{\it 2.~$\epsilon=1$, $\kappa>0$.}~The constraints \Ref{ineq1}, \Ref{ineq2} give
\bea
&&(A)\quad A_0\phi'^2_0<\frac{1}{4\pi\kappa}, \quad V_0>\frac{1}{8\pi\kappa},
\Non
&&(B)\quad A_0\phi'^2_0>\frac{1}{4\pi\kappa}, \quad \frac{1}{8\pi\kappa}<V_0<\frac12A_0\phi'^2_0.
\non
\eea
Thus, in the case $\kappa>0$, $\epsilon=1$ there are domains of initial values $A_0\phi'^2_0$ and $V_0$ which provide the flare-out conditions \Ref{flareout}. Stress also that the initial value $V_0$ should be necessarily positive at the throat.

\vskip6pt{\it 3.~$\epsilon=1$, $\kappa<0$.}~It is easy to check that Eqs. \Ref{ineq1} and \Ref{ineq2} are inconsistent in this case, and hence the flare-out conditions \Ref{flareout} are not fulfilled.

\vskip6pt{\it 4.~$\epsilon=-1$, $\kappa>0$.}~Eqs. \Ref{ineq1}, \Ref{ineq2} give
\beq\label{case4}
A_0\phi'^2_0<\frac{1}{4\pi\kappa}, \quad V_0>-\frac{1}{2}A_0\phi'^2_0.
\eeq

\vskip6pt{\it 5.~$\epsilon=-1$, $\kappa<0$.}~Eqs. \Ref{ineq1}, \Ref{ineq2} give
\beq\label{case5}
A_0\phi'^2_0>0, \quad -\frac{1}{2}A_0\phi'^2_0<V_0<\frac{1}{8\pi|\kappa|}.
\eeq
Thus, in case $\epsilon=-1$ there are domains of initial values $A_0\phi'^2_0$ and $V_0$ which provide the flare-out conditions \Ref{flareout}. It is worth noticing that the value $V_0$ is admissible, and so one may expect to get a wormhole solution without potential.

In the model with nonminimal derivative coupling there is a nontrivial case $\epsilon=0$, when the free kinetic term is absent. Let us consider also this case.
\vskip6pt{\it 6.~$\epsilon=0$, $\kappa>0$.}~Eqs. \Ref{ineq1}, \Ref{ineq2} give
$$
A_0\phi'^2_0<\frac{1}{4\pi\kappa}, \quad V_0>0.
$$
Thus, in this case there are domains of initial values $A_0\phi'^2_0$ and $V_0$ which provide the flare-out conditions. As well as for $\epsilon=1$ and $\kappa>0$, the initial value $V_0$ should be necessarily positive at the throat.

\vskip6pt{\it 7.~$\epsilon=0$, $\kappa<0$.}~Eqs. \Ref{ineq1}, \Ref{ineq2} are inconsistent.

\vskip6pt Summarizing, we can conclude that the flare-out conditions \Ref{flareout} in the model with nonminimal derivative coupling can be fulfilled for various values of $\epsilon$ and $\kappa$. Respectively, the flare-out conditions provide an existence of solutions with the throat. It is worth especial noticing that the throat in the model with nonminimal derivative coupling can exist not only if $\epsilon=-1$ (phantom case), but also if $\epsilon=1$ (normal case) and $\epsilon=0$ (no free kinetic term).

To finish the analysis of the field equations \Ref{fieldeq} at the throat, let us consider the metric function $A(r)$ and its first and second derivatives at $u=0$. The value $A_0$ is a free parameter. Though $A'_0$ is also free, we assume, just for simplicity, $A'_0=0$. Note that in this case $A(u)$ has an extremum at the throat $u=0$. Using Eq. \Ref{(00)-(22)}, we can find
\beq
A''_0=-\frac{8\pi}{r_0}\frac{\kappa A_0\phi'^2_0+2 r_0^2 V_0}{1-4\pi\kappa A_0\phi'^2_0}.
\eeq
The sign of $A''_0$ determines a kind of the  extremum of $A(u)$; it is a maximum if $A''_0<0$, and a minimum if $A''_0>0$. It is worth noticing that, with the physical point of view, the maximum (minimum) of $A(r)$ corresponds to maximum (minimum) of gravitational potential. In turn, the gravitational force equals to zero at extrema of the gravitational potential; moreover, in the vicinity of maximum (minimum) the gravitational force is repulsive (attractive). As a consequence, the throat is repulsive or attractive depending on the sign of $A''_0$.

As an example, let us consider the model with $\epsilon=-1$. By using the relations \Ref{case4} and \Ref{case5}, we can see that $A''_0<0$ if $\kappa>0$, and $A''_0>0$ if $\kappa<0$. Hence, the throat is repulsive in the first case, and attractive in the second one.

\subsection{Asymptotical analysis}
While the throat is an essential feature of the wormhole geometry, its asymptotical properties could be varied for different models. Traversable wormholes are usually assumed possessing two asymptotically flat regions connected by the throat, and in this paper we will look for wormhole solutions with an appropriate asymptotical behavior.

The spacetime with the metric \Ref{metric} has two asymptotically flat regions ${\cal R}_\pm:\ u\to\pm\infty$ provided $\lim_{u\to\pm\infty}\{r(u)/|u|\}=\delta_\pm$ and $\lim_{u\to\pm\infty}A(u)=A_{\pm}$. Since a flat spacetime is necessarily empty, we have also to suppose that
$\lim_{u\to\pm\infty}\phi(u)=\phi_{\pm}$ and $\lim_{u\to\pm\infty}V(\phi(u))=V(\phi_{\pm})=0$. Assume the following asymptotics at $|u|\to\infty$:
\bea
r(u)&=&\delta_\pm |u|\left[1+\frac{\alpha_\pm}{|u|}+O(u^{-2})\right],
\\
A(u)&=&A_\pm\left[1 -\frac{\beta_\pm}{|u|}+O(u^{-2})\right],
\\
\phi(u)&=&\phi_\pm\left[1 -\frac{\gamma_\pm}{|u|}+O(u^{-2})\right],
\\
V(\phi(u))&=&O(u^{-5}).
\eea
Substituting above expressions into Eq. \Ref{11} and collecting leading terms gives
\beq
A_\pm=\delta_\pm^{-2}.
\eeq
Thus, the asymptotical form of the metric \Ref{metric} is
\beq
ds^2=-\delta_\pm^{-2}\left(1-\frac{\beta_\pm}{|u|}\right)dt^2 +\delta_\pm^{2}\left(1-\frac{\beta_\pm}{|u|}\right)^{-1}du^2 +\delta_\pm^2 u^2\left(1+\frac{\alpha_\pm}{|u|}\right)^2 d\Omega^2.
\eeq
Introducing in the asymptotically flat regions ${\cal R}_\pm$ new radial coordinates
$$
\rho_\pm=\delta_\pm |u|\left(1+\frac{\alpha_\pm}{|u|}\right)
$$
and neglecting terms $O(\rho_\pm^{-2})$, we obtain
\beq
ds^2=-\left(1-\frac{2m_\pm}{\rho_\pm}\right)dt_\pm^2 +\left(1-\frac{2m_\pm}{\rho_\pm}\right)^{-1}d\rho_\pm^2 +\rho_\pm^2 d\Omega^2,
\eeq
where $2m_\pm=\delta_\pm\beta_\pm$ and $t_\pm=\delta_\pm^{-1} t$. This is nothing but two Schwarzschild asymptotics at $u\to\pm\infty$ with masses $m_\pm$. Taking into account that $\delta_\pm|u|=\lim_{u\to\pm\infty}r(u)$ and $\delta_\pm=\pm\lim_{u\to\pm\infty}r'(u)$ we can find the following asymptotical formula
\beq\label{m}
m_\pm=\lim_{u\to\pm\infty}[r(u)(1-r'^2(u)A(u))].
\eeq

\subsection{Exact wormhole solution with $\kappa=0$ }
Let us discuss the particular case $\kappa=0$ (no nonminimal derivative coupling). In this case the system of field equations \Ref{combfieldeq} reduces to well-known equations for a minimally coupled scalar field:
\begin{subequations}\label{system_k0}
\bea
&& \displaystyle
\frac{r''}{r}=-4\pi\epsilon\phi'^2,
\\
&& \displaystyle
(A'r^2)'=-16\pi r^2V,
\\
&& \displaystyle
A(r^2)''-A''r^2=2,
\\
&& \displaystyle
\epsilon(Ar^2\phi')'=r^2 V_{\phi}.
\eea
\end{subequations}
Supposing additionally $\epsilon=-1$ (phantom scalar) and $V=0$ (no potential term), one can find an exact wormhole solution to the system \Ref{system_k0} (see \cite{Ellis,Bro73}).
Adopting the result of \cite{SusZhang} we can write down the solution
as follows
\beq\label{staticmetric}
ds^2=-e^{2\lambda(r)}dt^2+e^{-2\lambda(r)}\left[du^2+(u^2+u_0^2)d\Omega^2\right],
\eeq
\beq\label{staticsf}
\phi(u)=\left(\frac{m^2+u_0^2}{4\pi m^2}\right)^{1/2}\lambda(u),
\eeq
where $\lambda(u)=({m}/{u_0})\arctan({u}/{u_0})$, and $m$, $u_0$ are two free parameters. Taking into account the following asymptotical behavior:
$$
e^{2\lambda(u)}=\exp\left(\pm\frac{\pi
m}{u_0}\right)\left[1-\frac{2m}{u}\right]+O(u^{-2})
$$
in the limit $u\to\pm\infty$, we may see that the spacetime with
the metric \Ref{staticmetric} possesses by two asymptotically flat
regions. These regions are connected by the throat whose radius
corresponds to the minimum of the radius of two-dimensional
sphere, $r^2(u)=e^{-2\lambda(u)}(u^2+u_0^2)$. The minimum of $r(u)$ is
achieved at $u=m$ and equal to
$$
r_{0}=\exp\left(-\frac{m}{u_0}\arctan\frac{m}{u_0}\right)(u^2+u_0^2)^{1/2}.
$$
Asymptotical masses, corresponding to $u\to\pm\infty$, are
$$
m_{\pm}=\pm m\exp(\pm\pi m/2u_0).
$$
It is worth noticing that the masses $m_\pm$ have both different values and different signs, and so wormholes supported by the minimally coupled scalar field inevitably have a negative mass in one of the asymptotical regions.

Note also that there is a particularly simple case $m=0$ when the static
solution \Ref{staticmetric}, \Ref{staticsf} reduces to
\beq\label{MTwh}
ds^2=-dt^2+du^2+(u^2+u_0^2)d\Omega^2,
\eeq
\beq
\phi(u)=(4\pi)^{-1/2}\arctan({u}/{u_0}).
\eeq
In this case both asymptotical masses are equal to zero, $m_\pm=0$, and so such the wormhole is massless.

\subsection{Numerical analysis}
In this section we present results of numerical analysis of the field equations \Ref{fieldeq}. Note that in order to realize a numerical analysis into practice one needs first to specify a form of the potential $V(\phi)$. The requirement of asymptotical flatness dictates $\lim_{u\to\pm\infty}V(\phi(u))=V(\phi_{\pm})=0$. The simplest choice obeying this asymptotical behavior corresponds to zero potential, and hereafter we will assume $V(\phi)\equiv 0$. As the initial condition analysis has shown, the flare-out conditions with $V_0=0$ are only fulfilled in case $\epsilon=-1$.

An initial conditions for the system of second order ordinary differential equations \Ref{fieldeq} read $u=0$, $r(0)=r_0$, $r'(0)=r'_0$, $A(0)=A_0$, $A'(0)=A'_0$, $\phi(0)=\phi_0$, $\phi'(0)=\phi'_0$. Without loss of generality one can set $A_0=1$ and $\phi_0=0$. Since $u=0$ is assumed to be a wormhole throat, one get $r'_0=0$. Now the throat's radius $r_0$ given by Eq. \Ref{r0} can be found as
\beq
r_0=\sqrt{\frac{1-4\pi\kappa\phi'^2_0}{4\pi\phi'^2_0}}.
\eeq
Then the only two free parameters determining the initial conditions remain: $A'_0$ and $\phi'_0$.

\begin{figure}
\begin{center}
\includegraphics[width=5cm]{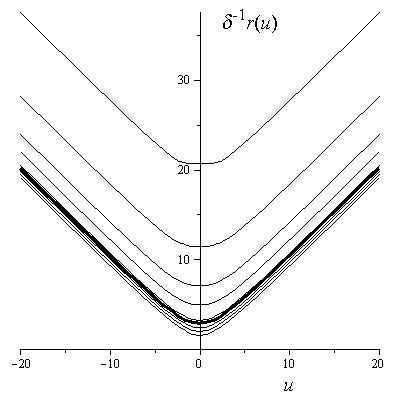} \includegraphics[width=5cm]{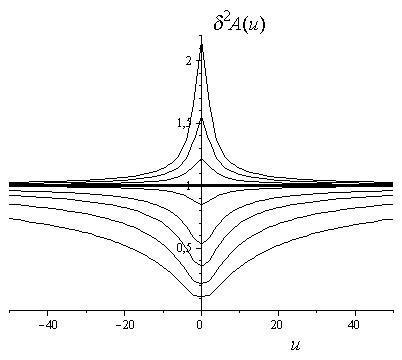}
\includegraphics[width=5cm]{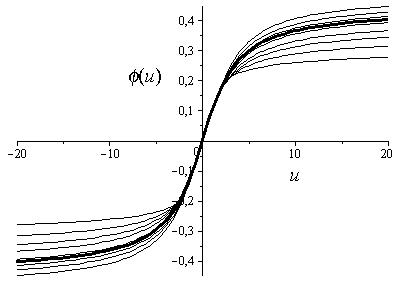}
\end{center}
\caption{Graphs for $\delta^{-1}r(u)$, $\delta^{2}A(u)$, and $\phi(u)$ are constructed for the initial values $A'_0=0$, $\phi'_0=0.1$ and $\kappa=-40,-20,-10,-5,-1,0,1,2,3$ (up-bottom for $r(u)$, bottom-up for $A(u)$, bottom-up for the right branch of $\phi(u)$); $\delta=\delta_+=\lim_{u\to\infty}r(u)/u$ ($\delta_+=\delta_-$). The bold lines correspond to $\kappa=0$ (no nonminimal derivative coupling).
\label{fig_rAP0}}
\end{figure}

\begin{figure}
\begin{center}
\includegraphics[width=6cm]{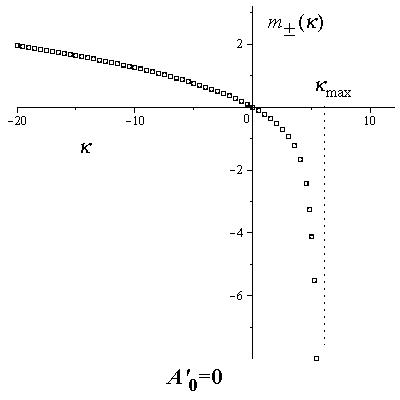} \includegraphics[width=6cm]{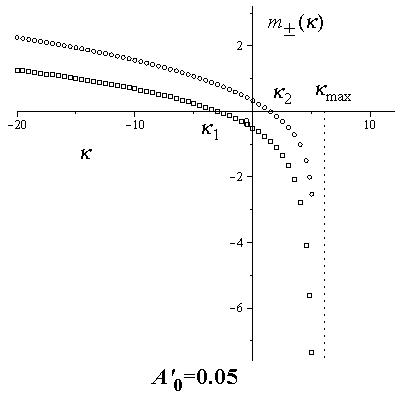}
\end{center}
\caption{Asymptotical masses $m_\pm$ vs $\kappa$ for $A'_0=0$ and $A'_0=0,05$. Note that $m_-=m_+$ if $A'_0=0$. In case $A'_0=0.05$ the masses $m_+$ and $m_-$ are represented by upper and lower graphs, respectively; $m_-(k_1)=m_+(k_2)=0$.
\label{fig_m}}
\end{figure}

First let us consider the case $A'_0=0$. In Fig. \ref{fig_rAP0} we represent numerical solutions
for $r(u)$, $A(u)$, and $\phi(u)$ for 
various values of $\kappa$. Note that both $r(u)$ and $A(u)$ are even functions possessing an extremum at the throat $u=0$; $r(u)$ has a minimum due to the flare-out conditions, and, as was mentioned above, $A(u)$ has a maximum if $\kappa>0$, and a minimum if $\kappa<0$. In case $\kappa=0$ one get $A(u)=1$; this coincides with the analytical result \Ref{MTwh}. The function $\phi(u)$ is odd; it smoothly varies between two asymptotical values $-\phi_+$ and $\phi_+$, where $\phi_+=\lim_{u\to\infty}\phi(u)$. The functions $r(u)$ and $A(u)$ given in Fig. \ref{fig_rAP0} also possess a proper asymptotical behavior: $\lim_{u\to\pm\infty}r(u)=\delta_\pm|u|$ and $\lim_{u\to\pm\infty}A(u)=\delta_\pm^{-2}$. In case $A'_0=0$ we have $\delta_-=\delta_+=\delta$ and the value of $\delta$ depends generally on $\kappa$, i.e., $\delta=\delta(\kappa)$. The asymptotical Schwarzschild masses $m_\pm$ corresponding to the numerical solutions $r(u)$ and $A(u)$ are shown in Fig. \ref{fig_m}. Because of the symmetry $u\leftrightarrow -u$ we have $m_-=m_+=m$. Moreover, $m$ is positive if $\kappa<0$, negative if $\kappa>0$, and $m=0$ for $\kappa=0$. Note also that wormhole solutions exist only for $\kappa<\kappa_{max}$, and $m\to-\infty$ if $\kappa\to\kappa_{max}$.

\begin{figure}
\begin{center}
\includegraphics[width=5cm]{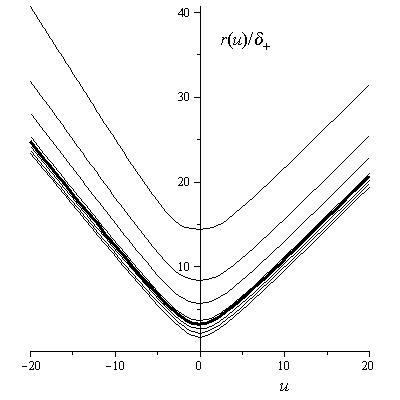} \includegraphics[width=5cm]{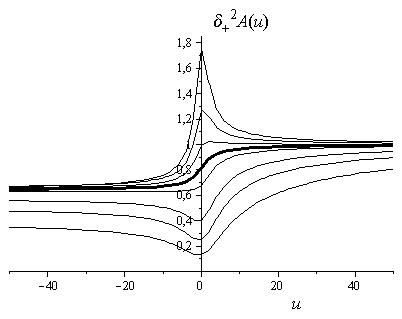}
\includegraphics[width=5cm]{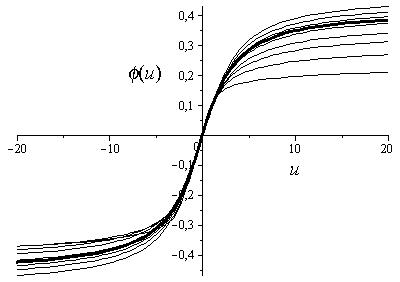}
\end{center}
\caption{Graphs for $\delta^{-1}r(u)$, $\delta^{2}A(u)$, and $\phi(u)$ are constructed for the initial values $A'_0=0.05$, $\phi'_0=0.1$ and $\kappa=-40,-20,-10,-5,-1,0,1,2,3$ (up-bottom for $r(u)$, bottom-up for $A(u)$, bottom-up for the right branch of $\phi(u)$); $\delta=\delta_+=\lim_{u\to\infty}r(u)/u$ ($\delta_+\not=\delta_-$). The bold lines correspond to $\kappa=0$ (no nonminimal derivative coupling).
\label{fig_rAP05}}
\end{figure}

The numerical solutions for $r(u)$ and $A(u)$ in the case $A'_0\not=0$ are shown in Fig. \ref{fig_rAP05}. It is worth noticing that in this case both $r(u)$ and $A(u)$ have different asymptotics at $u\to\pm\infty$: $\lim_{u\to\pm\infty}r(u)=\delta_\pm|u|$ and $\lim_{u\to\pm\infty}A(u)=\delta_\pm^{-2}$, where $\delta_-\not=\delta_+$. As a consequence, we get a wormhole configuration with two different asymptotical masses $m_\pm$. The value of $m_\pm$ depends on $\kappa$; this dependence is shown in Fig. \ref{fig_m}. Note that for $\kappa<\kappa_1$ both $m_+$ and $m_-$ are positive, for $\kappa>\kappa_2$ both $m_+$ and $m_-$ are negative, and for $\kappa_1<\kappa<\kappa_2$ (in particular, for $\kappa=0$) the asymptotical masses $m_\pm$ have opposite signs. Note also that $\kappa<\kappa_{max}$, and $m_\pm\to-\infty$ if $\kappa\to\kappa_{max}$.

Note that a qualitative behavior of numerical solutions for $r(u)$, $A(u)$, and $\phi(u)$ does not depend on the initial value $\phi'_0$, and above results were obtained for the specific choice $\phi'_0=0.1$.

\section{Summary and discussion}
We have constructed static spherically symmetric wormhole configurations in a gravitational theory of a scalar field with nonminimal derivative coupling to the curvature. Carrying out a local analysis of the  flare-out conditions, we have shown that solutions with a throat in the theory \Ref{action} with a nonzero derivative coupling parameter $\kappa\not=0$ could exist both if $\epsilon=-1$ and $\epsilon=+1$. For comparison, it is worth noticing that solutions with a throat in the minimally coupling model with $\kappa=0$ are forbidden for an ordinary scalar field with $\epsilon=+1$ \cite{Bro01}. Assuming additionally an asymptotical flatness, we have found numerical solutions describing traversable (Lorentzian) wormholes in the model with an arbitrary $\kappa$, $\epsilon=-1$ and $V(\phi)=0$ (no potential).

The wormhole solutions constructed in the paper could be classified by their asymptotic behavior which, in turn, is determined by asymptotics of the metric functions $r(u)$ and $A(u)$ at $u\to\pm\infty$. If $A'_0=0$, then both $r(u)$ and $A(u)$ are even, and so a wormhole configuration is symmetrical relative to the throat $u=0$. In this case both asymptotical masses $m_\pm$ are equal, i.e., $m_-=m_+=m$, and one has the following cases: (i) $m>0$ if $\kappa<0$; (ii) $m=0$ if $\kappa=0$; (iii) $m<0$ if $\kappa>0$. In case $A'_0\not=0$ a wormhole configuration has no symmetry relative to the throat, hence asymptotical masses $m_\pm$ are different, i.e., $m_-\not=m_+$, and one get (i) $m_->0$, $m_+>0$ if $\kappa<\kappa_1<0$; (ii) $m_-\le0$, $m_+>0$ if $\kappa_1<\kappa<\kappa_2$; (iii) $m_-<0$, $m_+\le0$ if $\kappa>\kappa_2>0$. Thus, depending on $\kappa$, a wormhole could possess positive and/or negative asymptotical Schwarzschild masses. It is worth emphasizing that both masses are positive only provided $\kappa<\kappa_1$ (for a symmetrical wormhole configuration one has $\kappa_1=\kappa_2=0$), otherwise one or both wormhole masses are negative. For example, let us consider the case $\kappa=0$, when the scalar field is minimally coupled to the curvature. In this case well-known wormhole solutions have been analytically obtained by Ellis \cite{Ellis} and Bronnikov \cite{Bro73} (see the discussion in Sec.~\ref{Sec}). Such the wormholes inevitably possess a negative Schwarzschild mass in one of the asymptotical regions, or, in the particular case of a symmetric wormhole configuration, both asymptotical masses are equal to zero. (Note that our numerical calculations given in Figs.~\ref{fig_rAP0}-\ref{fig_m} completely reproduce this particular analytical result.)

The stability of wormhole configurations is an important test of their possible viability. The stability of wormholes supported by phantom scalar fields was intensively investigated in the literature \cite{SusZhang,shin02,gon08,Bro_etal_11}, and the final resolution states that both static and non-static (see Ref. \cite{SusZhang}) scalar wormholes are unstable. Though this result is technically complicated, there is a simple qualitative explanation of this instability. Actually, as was mentioned above, a scalar wormhole inevitably possesses a negative Schwarzschild mass in one of the asymptotical regions; for example, let it be ${\cal R}_-:u\to-\infty$. This means that the gravitational potential is {\it decreasing} and the gravitational force is {\it repulsive} far from the throat. Consider now a small scalar perturbation of the wormhole geometry localized near the throat. Such the perturbation shall play a role of a small bunch of energy density and, because of the repulsive character of the gravitational field, it shall be pushed to the infinity ${\cal R}_-$. Similarly, any scalar perturbations will propagate from the throat vicinity to infinity, and this indicates an instability of the throat.

In comparison with wormholes supported by a phantom scalar field minimally coupled to the curvature, the scalar wormholes with nonminimal derivative coupling obtained in this paper have a more general asymptotic behavior. Namely, depending on a value of the nonminimal derivative coupling parameter $\kappa$ one of the following qualitatively different cases is realized: (i) one of the asymptotic wormhole masses or both of them are negative; (ii) both asymptotic masses are positive. Taking into account the previous qualitative consideration, one can expect that a wormhole configuration will be unstable in the first case and stable in the second one. Actually, if both asymptotic masses are positive, then the gravitational potential is {\it increasing} and the gravitational field is {\it attractive} on both sides of the wormhole throat. In this case all scalar perturbations should be localized in the vicinity of the throat, and this would provide a stability.\footnote{Note that wormhole solutions with positive asymptotical masses are known in the literature. Such wormholes were found as exact solutions of the scalar-tensor theory with nonminimal coupling $\xi R\phi^2$ and the $\phi^4$-potential \cite{SusKim} and the nonminimal Einstein-Yang-Mills model \cite{BalSusZay}. Also various thin-shell configurations represent wormholes with positive asymptotical masses; it is worth noticing that thin-shell wormholes could be stable (see, for example, Ref. \cite{Eir} and references therein).}

Of course, it is necessary to emphasize that the above consideration of wormhole stability has only a qualitative character. To answer finally the question -- are scalar wormholes with nonminimal derivative coupling stable or not? -- we need additional investigations which are in progress.


\section*{Acknowledgments}
The work was supported in part by the Russian Foundation for Basic Research grants No. 11-02-01162. S.S. appreciates Douglas Singleton and California State University Fresno for hospitality during the Fulbright scholarship visit.

\end{document}